\documentclass[aps,final,showpacs,showkeys,twocolumn,amssymb,groupedaddress]{revtex4-1}

\usepackage{amssymb}
\usepackage{amsmath}
\usepackage{showkeys}
\usepackage[english]{babel}

\begin{document}

\title{Swift J164449.3+573451 event: generation in the collapsing star cluster?}
\author{V.I. Dokuchaev}\thanks{e-mail: dokuchaev@lngs.infn.it}
\author{Yu.N. Eroshenko}\thanks{e-mail: eroshenko@inr.ac.ru}
\affiliation{Institute for Nuclear Research of the Russian Academy of
Sciences \\ 60th October Anniversary Prospect 7a, 117312 Moscow, Russia}

\date{\today}

\begin{abstract}
We discuss the multiband energy release in a model of a collapsing galactic nucleus, and we
try to interpret the unique super-long cosmic gamma-ray event Swift J164449.3+573451 (GRB 110328A
by early classification) in this scenario. Neutron stars and stellar-mass black holes can form evolutionary a
compact self-gravitating subsystem in the galactic center. Collisions and merges of these stellar remnants
during an avalanche contraction and collapse of the cluster core can produce powerful events in different
bands due to several mechanisms. Collisions of neutron stars and stellar-mass black holes can generate
gamma-ray bursts (GRBs) similar to the ordinary models of short GRB origin. The bright peaks during the
first two days may also be a consequence of multiple matter supply (due to matter release in the collisions)
and accretion onto the forming supermassive black hole. Numerous smaller peaks and later quasi-steady
radiation can arise from gravitational lensing, late accretion of gas onto the supermassive black hole, and
from particle acceleration by shock waves. Even if this model will not reproduce exactly all the Swift
J164449.3+573451 properties in future observations, such collapses of galactic nuclei can be available
for detection in other events.
\end{abstract}


\maketitle

\section{Introduction}

On March 28, 2011, the Swift's Burst Alert Telescope detected the unusual super-long gamma-ray
event Swift J164449.3+573451 from the center of a star-forming galaxy at the redshift $z=0.3534\pm0.0002$ \cite{Levetal11}. The burst sustained huge activity for a few days, and it keeps a softer but steady activity for
a few months. On the first days, the event had an extremely variable light-curve with several high peaks,
numerous smaller spikes, and a quasi-continuum. The peak luminosity of this burst reached a level of $\sim5\times10^{48}$~erg~s$^{-1}$. It is doubtful that this event can be
attributed to the class of superlong (with a duration
$>500$~s) GRBs \cite{TikSte05}, instead it seems to be a completely
different phenomenon. 

As a possible interpretation the of Swift J164449.3+573451 the tidal disruption flare
model of \cite{Ree88} was adopted in \cite{AlmAng11,Bloetal11,Bur11,Zau11}. In this model a star was destructed tidally by the supermassive black hole (SMBH) with a mass of $\sim10^7M_{\odot}$ in the galactic centre, the debris was accreted onto the SMBH, and it supposed that a jet was produced in a direction almost precisely to
the Earth. The later configuration is similar to a small blazar \cite{Bloetalblazar11}. The accidental jet direction to the Earth in Swift J164449.3+573451 for producing the huge variable activity.  

Here we propose an alternative explanation, based on the model of gravitational collapse of a central
compact cluster and multiple mergers of neutron stars (NSs) and black holes (BHs) there. Dense star
clusters are a typical structural element of galactic nuclei. Compact subclusters of stellar remnants
can form in the course of evolution due to mass segregation—sinking of the most massive stars to
the center and their explosions as supernovae \cite{Col67,San70,QuiSha90}. We hereafter refer to this inner subsystem of
compact objects as a ``cluster''. The cluster evolves dynamically due to two-body relaxation: the central
density and velocity dispersion grow, fast objects ``evaporate'' from the central core, forming a relaxed
isothermal halo around the central core. The central part (core) of this cluster approaches a relativistic
stage, and the possibility of gravitational collapse is limited by relaxation time \cite{Dok91}. 

There are several possible mechanisms of gamma ray generation and energy release in other bands during
and after the core collapse:
 
1. Collisions of neutron stars and stellar-mass black holes can directly generate gamma-ray bursts,
as is the case in the ordinary model of short GRBs. Before the collision, two remnants are captured
into a close orbit, and jet formation in this system can proceed similar to the numerical calculations
of \cite{GRBsim}. The fastest variability time-scale of Swift J164449.3+573451 is $\sim100$~s. So the separate 
GRBs could explain the high peaks in the light-curve of Swift J164449.3+573451 only if there is some mechanism of the peaks broadening as compared with usual short GRBs having durations $\le2$~s. This can be due to
a high gas density in the collapsing galactic nucleus. The prolonged emission lasts
$\sim10^4$~s in the model of ordinary and compacts stars collisions \cite{BarKom11}, \cite{BarKom10}. In the case of NS-NS and NS-BH collisions the extended emission can be sustained for $\sim100$~s, as it was shown in the two jet model of \cite{BarPoz11}.
Possibly, direct gamma pulses were not observed because the Earth was not located in the collimated jet directions, in this case
the followingmechanisms can be responsible for peak generation.

2. Each collision of compact objects (NS-BH) provides matter release in the form of debris, fireballs
and jets. This is an alternative source of gas accretion in addition to star tidal disruption \cite{AlmAng11,Bloetal11,Bur11,Zau11}, but the matter of a neutron star produces denser and more
compact (initially) ejection. Collisional destructions of several neutron stars provide a larger power then a
single ordinary star. Just after the SMBH formation, several (from different collisions) mutually fighting
accretion disks or one steady disk (along the angular momentum of SMBH) can form. The bright peaks
in the Swift J164449.3+573451 light curve during the first two days may therefore be a consequence
of multiple episodes of violent gas accretion onto the forming SMBH. Each NS collision and destruction
is responsible for a separate peak in the light curve of Swift J164449.3+573451. Then the accretion disk
forms, and different radiative processes develop in a similar way to the ordinary scenario with tidal destruction
of a star. Therefore many properties of the tidal destruction model could be relevant to our model
with only minor modifications.
 
3. Numerous smaller peaks can arise due to gravitational lensing of the signals (from mergers
and accretion) of the forming SMBH and compact objects in the cluster. Later lensings are similar to
``mesolensings'' \cite{BarEzo97} and can produce numerous subpeaks in the light curve.

4. Smooth flashes and quasisteady radiation can arise from gas accretion onto the SMBH at the time
of its formation and rapid growth (for a few first days) and later.

5. The galactic nucleus as a whole will be source of IR radiation because of multiple fireball formation
inside the gas-dust envelope. The debris of disrupted normal stars form the envelope in which
the NS/BH cluster is submerged. Really, the X-ray spectrum of Swift J164449.3+573451 shows strong
absorption which can witness a wide envelope around the central engine. Nevertheless, to explain Swift
J164449.3+573451 one must suppose that this gas envelope is optically thin for gamma radiation. Acceleration
of particles by multiple shock fronts of the fireballs is also possible.

6. This model also predicts huge multiple bursts of gravitational waves \cite{QuiSha87}. Similar events will be
detectable in the near future by new gravitational wave telescopes.

The first of the effects listed above is similar to the GRB recurrence scenario which has been elaborated
in \cite{DokEroOze97,DokEroOze98,DokEro11}. This model associates the recurrent and
multiple GRBs with the final dissipative stage of central star clusters in galactic nuclei, which precedes
SMBH formation. Accidental gravitational captures produce short-lived pairs of compact objects. According
to the most popular scenarios, short GRBs (with a duration of $\leq$~2~s)are generated during coalescence
of two NSs or a NS and a BH in binary systems \cite{BNPP84,Pacz86}. It has been shown in recent calculations \cite{GRBsim} that mergers really produce short GRBs. Similar but multiple GRBs can be generated during the cluster’s core gravitational collapse just before the SMBH birth. This model could also explain the unusual double burst GRB 110709B \cite{Bin-Binetal12}.

\section{Evolution of compact star cluster}

According to observations, about 50-80\%  of galaxies with low and intermediate luminosities contain the compact central star clusters \cite{Feretal06}, as well as our Milky Way \cite{MasCap11}. Most massive stars of these star clusters 
inevitably sank down to the dynamical centre and exploded as supernova long ago at the initial epoch of the star cluster evolution. Therefore, the formation of the compact subsystems of NSs and stellar mass BHs must be a common process in the Universe \cite{Col67}, \cite{San70}, \cite{QuiSha90}.
Let us consider the cluster of NSs and BHs, formed at the time $t_i\ll
t_{0.35}$, where $t_{0.35}\simeq10$~Gyr corresponds to the red
shift $z=0.35$. The subscript ``i'' marks the quantities at the
moment $t_i$. During the dynamical evolution of the cluster's core its mass
$M_c$, number of compact objects $N_c\equiv M_c/m$, velocity dispersion $v$, 
virial radius $R_c=GN_c m/2v^2$ and the masses of separate objects $m$ are all
evolving. We use the homological approximation for the mean
properties of the core. In this approximation the behaviour of the
core obeys the system of equation, obtained in \cite{QuiSha87}. 
The global parameters of the core change due to objects dynamical evaporation from the core and due to energy loss
for gravitational radiation, while the rate of the above processes and the inner structure of the cluster are
determined by the two-body relaxation and mergers of the compact objects.
We rewrite the system of equations from \cite{QuiSha87} through the new variables $x=v^2/c^2$ in the following form:
\begin{equation}
\left\{
\begin{array}{l}
\dot x/x=(\alpha_2-\alpha_1)t_r^{-1}+7t_{\rm cap}^{-1}/5
\label{gom1} \\
\dot M/M=-\alpha_2t_r^{-1} \\
\dot m/m=t_{\rm cap}^{-1},
\end{array}
\right.
\end{equation}
where $t_{\rm cap}$ takes into account the dissipative processes of
mergers of compact objects and gravitational waves emission, $t_{\rm
cap}^{-1}=\sigma_{\rm cap} v/\sqrt2$, and $t_r$ is the two-body
relaxation time (characteristic time-scale for stellar cluster
dynamical evolution)
\begin{equation}
 \label{tr}
 t_{\rm r}=\left(\frac{2}{3}\right)^{\!1/2}\!\!\!
 \frac{v^3}{4\pi G^2m^2n\Lambda},
\end{equation}
where $\Lambda=\ln(0.4N)$ is the Coulomb logarithm and we put $\Lambda\simeq15$ hereafter, i.~e. $\Lambda\simeq const$ with logarithmic accuracy, $n$ is the compact objects number density. 
According to \cite{QuiSha89}, the capture cross-section for the two objects with masses $m_1$ and $m_2$ is
\begin{equation}
\sigma_{\rm cap}\approx\frac{6\pi G^2(m_1+m_2)^{10/7}m_1^{2/7}m_2^{2/7}}{2^{10/7}c^{10/7}v^{18/7}},  \label{sigm1m2}
\end{equation}
and we put $m_1=m_2=m$ in this Section. The  coefficients
$\alpha_1=8{.}72\times10^{-4}$ and $\alpha_2=1{.}24\times10^{-3}$
were obtained in the Fokker-Plank model \cite{QuiSha87}. This 
formalism describes the contraction of the clusters' core, while the evaporated
objects form the isothermal halo around the core with density profile $\rho\propto r^{-2}$. The
halo can be involved into the avalanche contraction after the core reaches a relativistic stage, as will be
discussed in the next section.

Most of the cluster lifetime is non-dissipative (one can neglect
the $t_{\rm cap}$ terms). From the solution of (\ref{gom1}) in this case
we have:
\begin{equation}
M_c=M_i[1-(t-t_i)/(\kappa t_{ri})]^{\nu_1}, R_c=R_i[1-(t-t_i)/(\kappa
t_{ri})]^{\nu_2},\label{mr}
\end{equation}
where $\nu_1=2\alpha_2/(7\alpha_2-3\alpha_1)$,
$\nu_2=2(2\alpha_2-\alpha_1)/(7\alpha_2-3\alpha_1)$,
$\kappa=2/(7\alpha_2-3\alpha_1)$. The quantity $t_e\equiv\kappa t_{r,i}\approx330t_{r,i}$ approximately equals to the duration of the
cluster evolution up to the time of the full evaporation or 
gravitation collapse into a SMBH, because the final dissipative stage is rather short. 

The solution of the exact Equations (\ref{gom1}) has the form
\begin{equation}
m(x)=m_i(x^{5/7}+x_{\rm dis}^{5/7})/(x_i^{5/7}+x_{\rm
dis}^{5/7}),\label{mtmi}
\end{equation}
\begin{equation}
M_c(x)=M_i\left[ \frac{x_i^{5/7}(x_{\rm
dis}^{5/7}+x^{5/7})}{x^{5/7}(x_{\rm dis}^{5/7}+x_i^{5/7})}
\right]^{\frac{7\alpha_2}{5(\alpha_2-\alpha_1)}}, \label{mbigt}
\end{equation}
where
\begin{equation}
x_{\rm dis}\equiv\left[10\Lambda(\alpha_2-\alpha_1)/(7\sqrt{3})\right]^{7/5}
\simeq 5.3\times10^{-4}.
\end{equation}
The cluster of NS had formed long before the dissipative stage. Near the
dissipative stage $x\gg x_i$. Putting $m=2m_i$ in (\ref{mtmi}), we
obtain $x=x_{\rm dis}$. Therefore, the dissipative stage begins at
$v/c\simeq0.023$, i.~e. $v\simeq6.9\times10^3$~km~s$^{-1}$, independently
from the initial parameters of the cluster. Duration of the
dissipative stage
\begin{equation}
\tau_{\rm dis}=t_{\rm cap}(x_{\rm dis}) \simeq 4
\left(\frac{M_c}{10^4M_{\odot}}\right)^{2}
\left(\frac{m}{2.8M_\odot}\right)^{-1}\mbox{yr}.
\end{equation}
Collapse occurs at $x\simeq x_f\simeq0.1$ \cite{QuiSha87}. If we require the collapse at $z=0.35$, i.e. $\kappa
t_{r,i}=t_{0.35}\simeq9.8\times10^9$~yr, we get
\begin{equation}
x_i=9\times10^{-7}\left(\frac{M_i}{10^7M_{\odot}}\right)^{4/3}\left(\frac{m_i}{1.4M_\odot}\right)^{-2/3},
\label{xiexpr}
\end{equation}
and it corresponds to $v_i\simeq280$~km~s$^{-1}$.
If we require additionally that at the beginning of the collapse the
core hasn't evaporated, $M_c(x_f)>2m_i$, then we have from
(\ref{mbigt}) and (\ref{xiexpr})
\begin{equation}
M_i>M_{\rm min}\simeq3\times10^7M_{\odot},\label{mmin}
\end{equation}
or $N_i>2\times10^7$ if $m_i=1.4M_{\odot}$. The boundary mass
$M_{\rm min}$ sets the minimum for the initial mass of the cluster
in the scenario under consideration. Mass
segregation accelerates the evolution as compared
with this homological model. Therefore, the real $M_{\rm min}$ may be somehow smaller. We put in the
following estimations $M_{\rm min}\sim 10^7M_{\odot}$. Note that
till the moment $x_f\simeq0.1$ the masses $m$ of BH have time to
grow up to $\sim m_i(x_f/x_{\rm dis})^{5/7}\sim42m_i$, according to
(\ref{mtmi}), although the validity of (\ref{mtmi}) is doubtful near
the moment of the collapse. The homological model gives a
useful qualitative description, but numerical results must be accepted with some caution. This picture of
the dynamical evolution is somehow complicated in the presence of an angular momentum, but it is not
important in the central relaxed part of the cluster. A serious modification is required for a cluster with
a pre-existing massive BH at the center. In this case the central BH grows slowly instead of a fast
avalanche collapse of the cluster.

In plenty of works, the dynamical evolution of globular clusters was considered
in the ``evaporation model'' (see, e.~g., \cite{Dok91}). This model supposes that an object leaves the core
with zero kinetic energy, so that the total energy of the evolving core is conserved $E_c\propto M_c^2/R_c\approx const$.
The rate of evaporation was calculated from the high-velocity tail of Maxwellian distribution. 
Let us compare the evaporation model with the considered above model of \cite{QuiSha87}.  In the evaporation model $\alpha_1=0$ and $\alpha_2=7{.}4\times10^{-3}$, and the duration of core evolution till the collapse $t_e\simeq2/(7\alpha_2)\approx38.6t_{r,i}$ is several times less in comparison with the above calculations. It means that the core more easily reaches the relativistic stage and its initial mass $M_i$ could be smaller. For this model $x_{\rm dis}\simeq3.5\times10^{-2}$ ($v\simeq0.19c$), $\tau_{\rm dis}\simeq3.6$~yr, $x_i\simeq2\times10^{-7}$ ($v_i\simeq140$~km~s$^{-1}$) and $M_{\rm min}\simeq3\times10^6M_{\odot}$.

One can roughly suppose that the epoch of collapses lasts
$t_0\simeq10$~Gyr, then the total rate of such collapses in the
observable Universe is estimated to be
\begin{equation}
\dot N_h\sim\frac{4\pi}{3}(ct_0)^3n_gt_0^{-1}\approx0.1
\left(\frac{n_g}{10^{-2}\mbox{Mpc}^{-3}}\right) \mbox{yr}^{-1},
\label{ratecoll}
\end{equation}
where $n_g$ is the number density of structured galaxies with
central clusters. It seems natural that the collapse predicted by this model could be seen for several
years of observations. Moreover, the cluster’s core just before the collapse can be a promising source of
multiple gravitational wave bursts \cite{QuiSha87}.

\section{Avalanche contraction and gravitational
collapse of the central core}
\label{avalsec}

Now we consider the final highly dissipative stage of the cluster evolution. We suppose that the central
cluster consists of stellar-mass BHs with an admixture of non-coalesced NSs. The BH masses in the
cluster grow through mergers \cite{QuiSha87}. Suppose that their masses
immediately before the core's collapse are $m_{\rm BH}\sim10m_i$, where $m_i\simeq1.4M_{\odot}$. In the central core $m_{\rm
BH}\sim42m_i$ according to the rough estimation of the previous section, but we put $m_{\rm BH}\sim10m_i$ as mean over the cluster. We put
also that the cluster has the total mass $M$ (mainly in the form of
BHs) and the NSs constitute  a fraction $f_{\rm NS}\ll1$ of $M$. The
dynamical evolution of clusters is accompanied by the secular growth
of the central velocity dispersion or, equivalently, by the growth of the
central gravitational potential. Just near the collapse stage the cluster
has a dense core with mass $M_c$, radius $R_c\simeq3R_{g,c}$ (last stable orbit), where
$R_{g,c}=2GM_c/c^2$, and velocity dispersion $\simeq0.3c$. All objects with
a small angular momentum will fall onto the core without returning to the outer part of the cluster.
The avalanche occurs due to the presence of quasielliptical orbits that connect different layers of compact
objects \cite{ZelPod65}. Objects from the loss-cone are captured by the central core, and the central region collapses into the SMBH. Numerical calculations \cite{ShaTeu86} shows that the several percent of the
total cluster mass will collapse onto the forming SMBH during the several
dynamical times $t_{\rm dyn}=R_{\rm cl}/v$ -- duration of the avalanche
contraction, where $R_{\rm cl}$ is the radius of the whole cluster (halo around the core).

In view of $f_{\rm NS}\ll1$, the NS-BH coalescences are more probable
than the NS-NS ones. At the mass ratio $m_{\rm BH}\sim10m_i$, the NS
does not fall into the BH as a whole but is disrupted by tidal
forces and produces a fireball and GRB. The rate of NS-BH
coalescences in the core is $\dot N_c=\sigma_{\rm cap}uN_{\rm
NS}n_{\rm BH}$, where the gravitational capture cross section for
objects with different mass is given by
(\ref{sigm1m2}), $N_{\rm NS}=f_{\rm NS}M_c/m$ is the number of NSs
in the core, and $n_{\rm BH}$ is the BH number density in the core.
For the particular parameters $M=10^7M_{\odot}$ and $v=0.05c$ we have $R_{\rm cl}\simeq10^{-4}$~pc and $t_{\rm dyn}\simeq2.3$~days. The number of NS-BH mergers in the core during this time is
\begin{equation}
\dot N_c\, t_{\rm dyn}\simeq7\times\left(\frac{f_{\rm
NS}}{2\times10^{-5}}\right)\left(\frac{M}{10^7M_{\odot}}\right)
\left(\frac{M_c}{5\times10^5M_{\odot}}\right)^{-1}.
\label{nsbhnumber}
\end{equation}
The rate of NS-BH collisions in the entire cluster before the core
collapse:
\begin{equation} \dot N\simeq2.5\times10^{-2}
\left(\frac{f_{\rm NS}}{2\times10^{-5}}\right)
\left(\frac{M}{10^7M_{\odot}}\right)^{-1}
\left(\frac{v}{0.05c}\right)^{31/7} \mbox{yr}^{-1}, \label{e3}
\end{equation}
i.e., the cluster doesn't show the fast recurrence before the
collapse. Multiple short GRBs can be generated in the process of
avalanche-like contraction at the cluster center \cite{ZelPod65},
\cite{ShaTeu86}. Therefore, we have the several NS-BH collisions
and, accordingly, several GRBs-like events over the time of several days. Similar
temporal characteristics were required in \cite{DokEro11} to explain
the multiple GRB recorded on October 27, 1996. 

In approach to the collapse moment $t_{\rm coll}$ the rate of compact objects collisions (expression like (\ref{e3})) rises abruptly, and the asymptotic solution of (\ref{gom1}) near $t_{\rm coll}$ gives $\dot N_c\propto[1-(t-t_{\rm coll})/\tau^*]^{-2}$,
where the characteristic time-scale is
\begin{equation}
\tau^*\simeq0.007\times\frac{R_{g,c}^2}{cr_gx^{31/14}}\simeq0.8\mbox{~days},
\end{equation}
where $r_g=2Gm_{\rm BH}/c^2$. Therefore the stage of frequent NS-BH collisions lasts $\sim$~one~day. It explains why the host galaxy of Swift J164449.3+573451 didn't show activity long before the event. The first activity, which could be associated with the beginning of the collapse and the fast collisions, was found in the Burst Alert Telescope's data at 3 days before the main event, and the flux of the precursor was $\sim 7\%$ of the first main peak \cite{Bur11}.  

Just as the central relativistic region had appeared it began to capture objects which flew through this region inside the ``loss-cone''. For the object at the radial distance $r$ from the core's centre the loss-cone is characterized by the angle $\phi\simeq b/r$, where $b^2\simeq R_{g,c}r_p(c/v)^2$ and the value of $r_p$ is the same as in the process of gravitational capture of two objects \cite{QuiSha89} with masses $m_1=m_{\rm BH}$ and $m_2=M_c\gg m_{\rm BH}$:
\begin{equation}
r_p^{7/2}\simeq\frac{85\pi G^{7/2}m_{\rm BH}M_c^{5/2}}{12\sqrt{2}c^5v^2}.
\end{equation}
For the calculations we use the isothermal density profile of the relaxed cluster $\rho(r)=M/(4\pi R_{\rm cl}r^2)$ till the core radius $R_c$ and the isotropic velocity distribution.
The number of BHs inside the ``loss-cone'' is estimated to be
\begin{equation}
N_{\rm cone}\simeq N\int\limits_{R_c}^{R_{\rm cl}}\frac{4\pi\rho(r)r^2dr}{M}\frac{\pi b^2}{4\pi r^2}\simeq 5.6\times N^{5/7}\frac{R_{g,c}^{5/7}R_{\rm cl}^{2/7}}{R_g}.
\end{equation}
For the fiducial parameters, used above, we have numerically $N_{\rm cone}\simeq3\times10^5$. Therefore, the mass inside the loss-cone is 
sufficient for our model. 

After the depletion of orbits inside the loss-cone during the several dynamical times the objects fall onto the SMBH only due to  the slow diffusion of their orbits through the loss-cone boundary with the rate $\dot N_{\rm dif}\simeq N/(\lambda t_r)$, where $t_r$ is given by (\ref{tr}) and $\lambda\simeq\ln(R_{\rm cl}/3R_{g,c})$. For the parameters in use the loss-cone is empty, because
$\dot N_{\rm dif}t_{\rm dyn}<N_{\rm con}$, and at the post-collapse stage the SMBH's mass grows only slowly with the rate $\dot N_{\rm dif}m_{\rm BH}$.

\section{Radiation from the collapsing nucleus}

Let us consider different ways of energy release during and after the avalanche contraction and
SMBH formation in a cluster of compact objects.

\subsection{Direct gamma-radiation}

NS-BH collisions can produce the short GRBs. If the gamma-radiation is collimated into the 
$\Omega_\gamma$ solid angle, the number of the observed GRBs will be $\Omega_\gamma/4\pi$ fraction of the number of
NS-BH collisions (\ref{nsbhnumber}). Let us denote the energy release at each merge by
$E_0=10^{52}E_{52}$~erg. Then the mean power of the source
during the active stage $t_{\rm dyn}$ in the case of isotropic
emission is
\begin{equation}
L_{\rm GRB}=E_0\dot
N_c\simeq2.3\times10^{47}E_{52}\left(\frac{f_{\rm
NS}}{10^{-4}}\right)\mbox{~erg~s}^{-1}.
\end{equation}
This value coincides with the average luminosity of Swift J164449.3+573451. For
the larger $f_{\rm NS}\simeq10^{-3}-10^{-2}$ the $\dot N_c$ is also greater, and one can achieve even the
observed peak luminosity $\sim5\times10^{48}$~erg~s$^{-1}$ of
Swift J164449.3+573451. Alternatively, the peak signal can come from the single
NS-NS or NS-BH merge events with beamed radiation \cite{GRBsim}. In
this case the high luminosity is the result of collimation. But one must require some mechanism of the
pulses broadening from the typical $\sim2$~s duration of the short GRBs till the minimum time-variability scale $\sim100$~s of the Swift J164449.3+573451. The mechanisms for the prolonged or extended emission after the first short bursts were discussed in \cite{BarKom11}, \cite{BarKom10}, \cite{BarPoz11}. But the light-curve of the Swift J164449.3+573451 has no short $\sim1$~s pulses. A simple explanation of their absence is that we are not at one of the jets directions, and therefore another
mechanism must be responsible for radiation with $\sim100$~s  variability.  

\subsection{Radiation from the debris accretion}

Collisions between NSs and BHs provide a release of matter which can be accreted as in the star tidal
disruption models \cite{AlmAng11,Bloetal11,Bur11,Zau11}. Ejections from NSs could
be more dense and compact, so they could provide a larger power as compared with a single ordinary star.
Tidal disruption of a non-ordinary star (a white dwarf) tidal disruption by the SMBH with mass $\le10^5M_{\odot}$ as a model of Swift~J1644+57 was proposed in \cite{KroPir11} and the short time-scales in the light curve were explained. Here we propose an even more extreme model with the NSs multiple destructions. The difference is that NSs destructed not by the tidal forces of central SMBH, but in the collisions with stellar mass BHs in the cluster. In this case the characteristic time-scales $\sim100$~s are related to the SMBH horizon light crossing time $2GM/c^3$ as in the model with the star tidal destruction.

Note that the orbits of NSs in the cluster are distributed almost isotropically. This means that
the debris from the collisions fall onto the central SMBH from different directions. If each collision produces
a separate short-lived accretion disk around the SMBH, the corresponding jets from different disks
will also be pointed isotropically. We can reproduce the configuration of a small blazar with an accidental
jet direction to the Earth in this case only by requiring that the actual number of collisions was larger then the number of the observed peaks by a factor of $4\pi/\Omega_j$, where $\Omega_j$ is the solid angle of the jet collimation. 

Another possibility states the existence of a steady accretion disk around the SMBH that supports a
single jet in the blazar configuration with a matter feed from the NS-BH collisions. In this model the
jet direction is not related to the orientation of the debris orbits, but can be determined, for example, by
the SMBHspin that had originated from the common rotation of the parent cluster. Radio and other signals
can be generated as in the model with star tidal destruction, and we will not repeat the corresponding
calculations here.

\subsection{Gravitational lensings}

Let us first consider the lensing of the point source on the point lens in the case the both are inside the same cluster at the distances from the Earth $D_s$ and $D_l$ respectively, so that $D_s\approx D_l\approx 1$~Gpc (galaxy at $z=0.35$), and $D_s-D_l$ is of the order of the cluster radius $R_{\rm cl}\simeq 10^{-4}$~pc. This configuration is similar to the  ``mesolensings'' in globular clusters \cite{BarEzo97}. When the Einstein radius of the lens is 
$$
R_E=\left(4GM_l/c^2\right)^{1/2}\left[D_l(1-D_l/D_s)\right]^{1/2}
$$ 
\begin{equation} 
\simeq4\times10^9\left(\frac{M_l}{10M_\odot}\right)^{1/2}\left(\frac{D_s-D_l}{3\times10^{14}\mbox{~cm}}\right)^{1/2}\mbox{~cm},
\end{equation}
where $M_l$ is the lens' mass. If one uses the $r^{-2}$ density profile of the cluster and suppose that the collision occurred in the core, then the integration gives the number of lensing BHs inside the Einstein radius at the level $N_l\sim5\times10^{-4}$. A similar small chance is in lensing on the central supermassive BH because the
large mass is compensated by a small distance from the source to the lens (collisions are most probable
in the core, near the central BH). This means that lensing of a point source or jet on BHs is unlikely.

The large fireball lenses easily, because the radius of the cluster $R_{\rm cl}\sim10^{-4}$~pc is of the order of the size $R_{\rm sw}$ of shock waves in GRBs where some part of the gamma-radiation are generated. First of all, the
fireball scattered on the forming central SMBH. The hot medium flew around the SMBH and was swallowed
inside it. Let the observer and an NS-NS or NS-BH merging be on opposite sides of the SMBH.
The surface of the fireball is not spherical from the observer’s side, it has a cavity in the SMBH direction.
As a result, for every small BH in the volume $\sim\pi R_{\rm sw}^2R_{\rm cl}$ there is amoment of fireball evolution when
the rays, normal to the fireball surface, go through the Einstein radius of the BH. Therefore multiple
lensing events are possible, like ``mesolensings'' in globular clusters \cite{BarEzo97}. But in contemporary models of
GRBs the main flux of gamma radiation comes from jets, therefore the fireball surface is a subdominant
source. In the case of A fireball scattered on the SMBH, different parts of its surface can be lensed
similarly to \cite{BabDok00}. These lensings can explain the observed fast variability of the light curve of Swift J164449.3+573451 during the first days. A characteristic feature of lensing is synchronous radiation in
different bands.

\subsection{Quasi-steady radiation}

Multiple NS and BH collisions first produce dense flows of debris and then an extended hot gas
cloud in the central region of the cluster. Hot gas from the surrounding cloud falls onto the SMBH
and radiates with the Eddington luminosity $L_{\rm E}=4\pi GM_{\rm SMBH} m_p c/\sigma_{\rm T}$, where $m_p$ is the proton mass and $\sigma_{\rm T}$ is the Thompson cross-section. This luminosity is $1.5\times10^{45}$~erg~s$^{-1}$ in the case of $10^7M_{\odot}$ SMBH which is the minimum inter-peak luminosity during the first several days of Swift
J164449.3+573451. Radiation from slow gas accretion onto the newly formed SMBH requires the gas
envelope to be thin for radiation.

\subsection{Multiple fireballs in the dense cloud}

The cluster of NSs and BHs is enveloped by a
gas cloud originating from destruction of ordinary
stars and partly from primordial gas in the galactic
center. Multiple GRBs are generated in the medium
``prepared'' by the preceding GRBs, so the evolution
of multiple fireballs must be considered as in  \cite{BerDokHidden}, \cite{BerDokHidden06}.

Collisions of compact objects produce the relativistic shock with
Lorentz factor $\Gamma\gg1$, which propagates ahead of the fireball
pushing the gas in the nucleus. The shock becomes non-relativistic
at the Sedov length $l_s =[3E_0/(4\pi \rho_e c^2)]^{1/3}$, where
$\rho_e$ is the density of the gas in the envelope. The length $l_s$
gives an estimate for cavity radius, produced by the single
fireball in the absence of the central SMBH. The multiple expanding fireballs in the cavity have the
shape of thin shells and are separated by the distance $R_c= ct_{\rm
cap}$, where $t_{\rm cap}$ corresponds only to NS-NS and NS-BH collisions in this case. The gas between each two fireballs is swept up and compressed by the preceding one. Multiple fireballs produce the stationary
cavity \cite{BerDokHidden} with radius $R_{\rm cav} =[\dot N_c
E_0/(4\pi \rho_e c_s^3)]^{1/2}$, where $c_s$ is the sound speed in
the gas. For the relativistic fireballs the gravitational attraction of the 
central SMBH is not important except for the later non-relativistic stage with $\Gamma\leq1$. 
The SMBH causes a deformation of the fireball, opening the conditions for
gravitational lensing. One can assume an equipartition
of the magnetic field induced by turbulence and by the dynamo mechanism. In this situation the
Fermi II acceleration mechanism operates, and high energy particle creation is possible. The signatures
of these processes are expected from the direction of Swift J164449.3+573451.

\section{Conclusion}

We propose a scenario of multiband powerful signal generation in an evolved star cluster in a galactic
nucleus during the gravitational collapse of its dense central core. This model naturally provides necessary
conditions for the event Swift J164449.3+573451 because the huge power, duration, and variabilitymay
indicate a more catastrophic event (or sequence of events) than tidal destruction of a single star.

In the typical case, the dynamical evolution of a central star cluster results in collisions and destruction
of a substantial fraction of stars and formation of a gas envelope \cite{Dok91}.
Simultaneously, a compact cluster of NSs and BHs can form in its evolution,
due to sinking of massive stars to the center of the cluster and their explosions as supernovae \cite{Col67,San70,QuiSha90}. The evolution of such a cluster leads to an avalanche contraction and gravitational collapse of the central
region \cite{QuiSha87}. This mechanism cannot explain the appearance of all SMBHs, especially in quasars at high redshifts, because of the lack of dynamical time. But star cluster collapses are nevertheless inevitable
in many galaxies at low redshifts. Recently, the sub-dominant role of galaxies’ mergers in triggering
the AGN activity was revealed \cite{Alletal11}. Avalanche contractions and collapses of star clusters could be
an alternative mechanism for AGN ignition at small redshifts along with tidal disruptions of stars or disk
instabilities.

Multiple dense gas releases and fireballs in NS-BH collisions can produce powerful peaks in the light
curve of Swift J164449.3+573451. The additional inter-peak radiation comes from gas accretion onto
the forming SMBH. We can expect that the properties of fireballs and gamma radiation in this model are
distorted by their scattering on the growing central SMBH (the characteristic time scale is $2GM/c^3\sim 100$~s), and by gravitational lensing on compact objects in the cluster.

\section*{Acknowledgments}

The authors thank S.B. Popov for fruitful discussions and the anonymous referee for useful comments
and suggestions. This study was supported in part by the grants OFN-17, RFBR 10-02-00635, and NSh-
871.2012.2.

\end{document}